# PHYSICAL BASIS OF SATRO ® – A NEW METHOD FOR ANALYSIS OF THE CARDIAC MUSCLE DEPOLARISATION

Jerzy Stanisław Janicki

*Medical Physics Research Institute in Poznan, Poland*

## Abstract

On the basis of the model of the current flow through a single fibre, changes in the electric charge density over the myocardium are described. With the use of relevant analytic formulae, supported with numerical calculations, the distribution and time dependencies of electric potentials on the surface of the thorax have been determined. The results obtained are compared with empirical data. A strong correlation between the theoretical predictions and the experimental data has been obtained. The model in question permits examination of instantaneous potentials resulting from electrical activation of particular segments of the cardiac muscle.

Keywords:

cardiac muscle ischemia, CM, depolarization, ECG, electric potentials, electrocardiogram, electrocardiographic signals, electrocardiography, heart, myocardium, QRS, SATRO

## Introduction

The paper reports results of a fragment of study realised at the Medical Physics Research Institute within the project on physical phenomena related to the cardiac muscle activity and numerical analysis of electric pulses generated by this activity.

Interpretation of the standard electrocardiography signals is usually based on the model introduced by Einthoven describing the electric field generated by the cardiac muscle in terms of a distribution of charges forming an electric dipole. A more complex structure of charges distributed in the cardiac muscle volume is proposed by the theory of multipoles [1] permitting analysis of the spatial distribution of potentials appearing around the cardiac muscle [2,3]. Another approach to the electric signals produced by the cardiac muscle is presented in a classical work of Noble [4] based on the Hodgkins-Huxley equations and analysis of potentials in the Purkinje fibre. The properties of the cells are described by a function of 23 variables describing 12 types of ionic flows in the cardiac muscle cells [5]. The flows related to the ion transportations (positive e.g. $Na^+$, $K^+$, $Ca^{++}$ or negative like $Cl^-$) are nonlinear in character [6-8] and the equations describing their dynamics can be solved only by complex numerical techniques [9]. Analysis of ECG signals is also often performed with the use of advanced mathematical methods [10-15].

Despite the common access to numerical methods it often happens that changes in the electric potential cannot be explained on the basis of the hitherto assumed theories. Still, often surprising observations are significant changes in the ECG picture of healthy patients and no changes in patients diagnosed with serious cardiological conditions.

The aim of the paper is to present a new model describing the appearance of transient potentials on depolarisation of the cardiac muscle chamber, taking into account the space-time relations and physiological and pathological factors, and indicate its advantages for evaluation of individual activations.

## 1. Physical model of the myocardium electric activity

The method of analysis of the electrocardiographic signals presented in this paper is based on a physical model of the appearance and propagation of an electric charge wave on depolarization of different segments of the ventricles. Each of the segments corresponds to a bundle of working conducting fibres along which a variable in time distribution of a resultant charge density appears. It is accompanied by the emergence of the instantaneous electric potential, whose value depends on physiological and pathological factors. The fibres investigated are built of cylindrical-shaped cells of about 15 μm in thickness and 100 μm in length, usually arranged parallel to each other with a definite number of branches and interconnections. Between the cells there is a direct contact through specific links of low resistance called *gap junctions* [18]. Propagation velocity of the charge wave along the fibres depends on the efficiency and the number of the junctions, which are considerably more numerous at the ends than on the lateral fibre edges [19]. Additionally, common membrane parts of adjacent cells ensure a free flow of ions and small hydrophilic molecules between subsequent cells, which prompts fast transfer of the activation along the fibre. The main role in this process is played by the ion channels in extra- and intracellular biological membranes [20]. The intracellular membranes make 90% of the cell membranes, and mitochondric ion channels have an important function in the cell ionic homeostasis [21, 22]. As a result of the subsequent activation and deactivation processes of the channels, there is a local increase in the density of ions taking part in the transport of the electric charge. Thus, the areas with the charges are not static, and the dynamics of their distribution resembles the propagation of a charge wave along a given fibre of the myocardium. [23]. A group of such fibres activated simultaneously in one of *CM* segments forms a bundle along which a wave of a resultant charge moves. The electric charges in individual bundles are activated at different moments in time, contributing independently to the total electric potential. In effect we can observe the appearance of potentials undergoing significant changes in position and time. Consequently, the multi-centred process of instantaneous excitations [16] reflects the depolarisation of particular *CM* regions, giving rise to the appearance of anisotropic potential distributions around the myocardium. The charge changing on *CM* depolarization triggers a flow of current [17], which is also a source of anisotropic magnetic field distribution. Time dependencies and distributions of magnetic fields for each of the instantaneous excitations will be described in detail in a separate paper.

Actually, the cardiac muscle is built of the three characteristic layers composed of the subendocardial (*W*), M type (*M*) and subepicardial (*N*) cells [24,25]. The model proposed describes the formation and movement of a resultant charge wave along the *CM* fibre in the extra- (I) and intracellular (II) regions. The waves of electric charge density are generated by the flow of sodium ($Na^+$), potassium ($K^+$) and chloride ($Cl^-$) ions into the cells of the bundle.

The rest state of the cells is a result of both active transport of ions $Na^+$ and $K^+$ and diffusive and electric transport of ions $Na^+$, $K^+$ and $Cl^-$. It should be remembered that also other ions, like calcium ion $Ca^{2+}$, take part in the process. However, to preserve clarity of the model, which describes changes in distribution of the electric charge density on depolarization, these ions are not shown.

Actually, each of the areas (I and II) in its volume is electrically neutral, however, in the area of a thickness of about 1μm around the cellular membrane there is a distinct layer of negative charges inside the cell and a layer of positive charges outside it. This insignificant surplus of heteropolar charges makes a difference in potentials of a value of about -$90 \times 10^{-3} V$ between the areas (I and II) in which there is different concentration of ions of the same kind. The density of $Na^+$ and $Cl^-$ ions is much higher in the extracellular area, whereas that of potassium ions $K^+$ and organic anions $A^-$ is much higher inside the cells. The organic anions are the negatively charged amino acids and proteins remaining practically continuously outside the cells. The above analysis of the ion distribution in the extra and intracellular areas concerns the cells in the rest state. This state changes when an electric pulse from the stimuli conducting system reaches Purikinje's junctions placed on the side of the endocardium. Then the potential between the areas increases, forcing a growing flow of sodium ions ($Na^+$) into the cells. Almost simultaneously the cells in deeper layers of the ventricle walls are activated triggering a similar effect in the second (*M*) and the third (*N*) part of the fibre. In each of these parts, independent changes in a resultant electric charge density take place, which substantially influence the total charge produced by the whole fibre. As a result of the process we observe the emergence of regions of a greater concentration of negative charge in area I and those of a greater concentration of positive charge in area II. The charge moves along the fibre in the direction of propagation of the activation. After the passage of the wave front, selective extra- and intracellular ion channels are activated, enabling the flow of ions both between areas I and II as well as inside the activated fibre cells. Because of substantially higher conductivity of area II, the propagation of electric charge takes place mainly inside the excited fibre in the direction opposite to that of depolarization. Consequently, each *CM* working fibre can be treated as a sort of linear conducting system, stretched between the endocardium and epicardium, through which the current related to ion transport flows.

## 2. Electric charge density

At the beginning we consider changes in the electric charge density triggered by the ion transport on the propagation of the depolarisation wave along one part of the fibre. In the model proposed, an increase in the flow of $Na^+$ ions into the cells causes a change in the positive $\rho_+$ and negative $\rho_-$ electric charge density in the intra- (II) and extracellular (I) areas, respectively. Each of these electric charge density values is determined by the charge appearing in the volume element $\Delta v'$ in the time $t'$ and $t'+\Delta t'$. Denoting the charge density connected with the front of the moving wave in time $t'$ as $\rho_{t'\pm}$ and the charge density corresponding to the time $t'+\Delta t'$ (related to conducting properties of a particular part of the bundle) as $\rho_{(t'+\Delta t')\pm}$, we obtain the following relation:

$$\rho_\pm = \rho_{t'\pm} + \rho_{(t'+\Delta t')\pm} \qquad (1)$$

As the areas I and II are characterised by quite different conducting properties, we can consider a resultant charge density obtained taking into regard both positive and negative charges. So, we introduce the resultant charge density $\Delta\rho$ defined with the formula:

$$\Delta\rho = \rho_+ + \rho_- \qquad (2)$$

One should note that the density $\rho_-$ is connected with the negative charge, thus in specific calculations we take the values of the appropriate parameters with a minus sign.

Propagation of the charge waves can be described by means of a commonly applied function related to the probability of activation and deactivation of the membrane [17]. It is a function of the following type:

$$f(x) = \frac{A}{1 + exp(x/k)} \qquad (3)$$

Because of the dynamics of the ion flows, the resultant charge density $\Delta\rho$ depends on both time ($t'$) and position ($x'$) of the volume element $dv'$ in the fibre studied. This is why the function was modified in such a way so that it would reflect both dependencies. We assume that the coordinate $x'$ is related to the direction of the fibre. Then changes in the charge density along its axis have a decisive effect on the $\rho_\pm$ value. Consequently, the changes in $\rho_{t'\pm}$ can be described by the function:

$$\rho_{t'\pm}(x',t') = \pm A \frac{1}{1 + exp(-(t'-\Delta t')/k)} \cdot \frac{1}{1 + exp(-(x'-\Delta x')/k)} \qquad (4)$$

where A is a constant parameter, whose value determines the amplitude of the wave front and depends on the number of conducting fibres in a given *CM* bundle. The parameter *k* is related to the propagation velocity of the disturbance along the fibre that is to the rate of the amplitude increase. The charge density $\rho_{(t'+\Delta t')\pm}$, can be expressed as:

$$\rho_{(t'+\Delta t')\pm}(x',t') = \frac{\pm a(x',t')}{1 + exp((x'+\Delta x')/k_{x'\pm} - (t'\pm\Delta t')/k_{t'\pm})} . \qquad (5)$$

The parameters $k_{x',t'}$ occurring here are connected with the propagation velocity of the charge waves in each of the regions of a given part of the fibre studied. Changes in the value of the expression $a(x',t')$ can be described by the function:

$$a(x',t') = A_G \left( \frac{1}{1+exp(-(t'-\Delta t')/k)} - \frac{1}{1+exp(-(t'+\Delta t')/k)} \right) \cdot \left( \frac{1}{1+exp(-(x'-\Delta x')/k)} - \frac{1}{1+exp(-(x'+\Delta x')/k)} \right) \tag{6}$$

The parameter $A_G$ defines a characteristic amplitude of the charge density for a given *CM* bundle after the wave front has passed. Due to the fact that during the wave front formation much more ions flow in an orderly manner than in the time $t'+\Delta t'$, the parameters characterising the amplitudes meet the condition $|A| >> |A_G|$. As follows from the above dependences, the highest value of the resultant charge density $\Delta \rho$ is observed in the area of a geometric centre of a given fibre part (*W,M,N*), and it is reached in the middle of the process duration. Thus, the changes in the positive and negative charge density in areas I and II finally take the form:

$$\rho_\pm(x',t') = A_G \left( \frac{1}{1+exp(-(t'-\Delta t')/k)} - \frac{1}{1+exp(-(t'+\Delta t')/k)} \right) \cdot \left( \frac{1}{1+exp(-(x'-\Delta x')/k)} - \frac{1}{1+exp(-(x'+\Delta x')/k)} \right) \cdot \frac{1}{1+exp((x'+\Delta x')/k_{x'\pm} - (t'\pm\Delta t')/k_{t'\pm})} \pm A \frac{1}{1+exp(-(t'-\Delta t')/k)} \cdot \frac{1}{1+exp(-(x'-\Delta x')/k)} \tag{7}$$

Graphic presentation of this dependence for one of the fibre parts is given in *Fig. 1*, where $\rho_+$ and $\rho_-$ describe changes in the electric charge outside and inside the fibres, respectively. The calculated distribution of the resultant charge density given by eq. (2) is shown in *Fig. 2*.

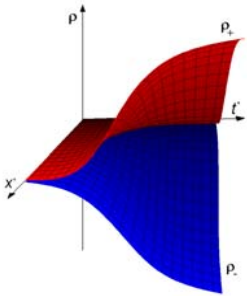

***Fig.1.*** *Change in electric charge density in the intracellular $\rho_+$ and extracellular area $\rho_-$ along one of the fibre parts (x') on the fibre depolarisation (t').*

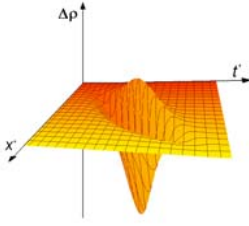

*Fig.2.* Change in the resultant of electric charge density $\Delta\rho$ along one of the fibre parts (x') on the depolarization process ( t') .

As it has been indicated above, the *CM* fibre structure is more complex, therefore for full description of the charge density distribution in the whole fibre of a length $L = L_W + L_M + L_N$ one should consider changes in each of its parts (*W,M,N*) of the lengths $L_W, L_M, L_N$, as shown in *Fig. 3*. In further analysis it will be more convenient to describe the boundary points in the coordinates: $x_0, x_W, x_M, x_N$.

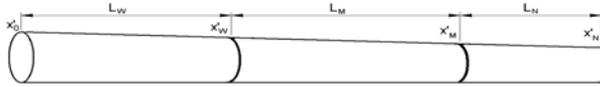

*Fig.3.* A method of determination of the lengths $L_W, L_M, L_N$ and coordinates $x_0, x_W, x_M, x_N$ for individual parts of the CM fibre (W,M,N).

Assuming that $t'$ is the time interval in which the depolarisation wave flows along the whole fibre and that $x'_j$ is within the range $\langle -L_j/2, L_j/2 \rangle$ for $j \in \{W, M, N\}$, then the charge density of the whole fibre can be determined by the values $(\rho_W, \rho_M, \rho_N)$ corresponding to its particular parts:

$$\rho_{\pm}(x', t') = \begin{cases} \rho_W(x'_W, t'_W) & dla \quad x'_O \leq x' \leq x'_W \\ \rho_M(x'_M, t'_M) & dla \quad x'_W < x' \leq x'_M \\ \rho_N(x'_N, t'_N) & dla \quad x'_M < x' \leq x'_N \end{cases} \quad (8)$$

Changes in the resultant charge density starting at different time moments in particular parts of the fibre bundle produce the variable in time electric charge.

## 3. Electric charge and dipole moment

In the model discussed the change in the characteristic quantities describing a given charge density wave, induces changes in the electric parameters of the myocardium area of interest. The total charge in a given area as well as moments of different orders, including the electric dipole moment, can change.

The spatial distribution and time changes in the total electric charge on depolarization of a given fibre are related to a distribution of positive $\rho_+$ and negative $\rho_-$ charge density in the following way:

$$q(t') = \int_F (\rho_+(\vec{r}',t') + \rho_-(\vec{r}',t'))d\vec{r}' \qquad (9)$$

where the integration is carried out over the area of the whole fibre $F$, and the vector $\vec{r}'$ describes the fibre position in a given coordinate system. One should remember that in the model proposed the charge densities are linear (they give charge per length unit). The waves of the negative and positive charges moving in the same direction have a similar character; however, the key issue is their interrelations. According to the above analysis, such parameters as the amplitudes of these waves, their mutual positions and the velocity of propagation of the disturbance in each of the fibre studied are of great significance. Depending on the myocardium condition the values of the above parameters can change.

In a specific case, the resultant charge density from a given bundle can be zero. Then the amplitudes of both waves are the same, and so are the velocities of the ions distribution. However, the cardiac muscle activation is always accompanied with a time-dependent electric charge. Moreover, the final effect can also depend on a dipole moment occurring in the myocardium fibres, whose vector $\vec{p}$ [26] is presented in the form:

$$\vec{p}(t') = \int_F \vec{r}'(\rho_+(\vec{r}',t') + \rho_-(\vec{r}',t'))d\vec{r}' \qquad (10)$$

The influence of higher order moments on the value of the potential measured is insignificant, therefore, their participation was omitted in further analysis. Due to the complex character of functions describing the charge density changes, analytical calculation of the total charge or dipole moments is practically impossible. Nonetheless, regarding the geometry of a single fibre it can be assumed that we have to do with a one-dimensional charge distribution and in effect the fibre studied can be treated as a one-dimensional structure. Then vector $\vec{r}'$ in equations (9 and 10) is replaced by the coordinate $x'$, which largely simplifies calculations, and the time changes in the electric charge and the dipole moment are described by the equations:

$$q(t') = \frac{A_+}{k_{t'}} \sum_{j=1,2}(-1)^j \left[ k_{x'}k_{t'} \ln\left( e^{-\frac{x'_j k_{t'}+k_{x'}(t'+dt')}{k_{x'}k_{t'}}} +1 \right) + x'_j k_{t'} + k_{x'}(t'+dt') \right]$$
$$- \frac{A_-}{k_{t'}} \sum_{j=1,2}(-1)^j \left[ k_{x'}k_{t'} \ln\left( e^{-\frac{x'_j k_{t'}+k_{x'}(t'-dt')}{k_{x'}k_{t'}}} +1 \right) + x'_j k_{t'} + k_{x'}(t'-dt') \right] \quad (11)$$

$$p(t') = A_+ \frac{k_{x'}}{k_{t'}} \sum_{j=1,2}(-1)^j \left[ k_{x'}k_j\, dilog\left( e^{-\frac{x'_j k_t+k_x(t'+dt')}{k_x k_t}} +1 \right) + x'_j k_{t'} \ln\left( e^{-\frac{x'_j k_t+k_x(t'+dt')}{k_x k_t}} +1 \right) \right]$$
$$- A_- \frac{k_{x'}}{k_{t'}} \sum_{j=1,2} \left[ k_{x'}k_{t'}\, dilog\left( e^{-\frac{x'_j k_t+k_x(t'-dt')}{k_x k_t}} +1 \right) - k_{t'} x'_j \ln\left( e^{-\frac{x'_j k_t+k_x t'}{k_x k_t}} + e^{\frac{dt'}{k_t}} \right) + x'_j dt' \right] \quad (12)$$

where a special dilog $(x')$ is defined in [27] by the relation:

$$dilog(x') = \int_1^{x'} \frac{\ln(\tau)}{1-\tau} d\tau \quad (13)$$

Although only one fibre part and a partial change in the charge density have been taken into account in the above analysis, the formulae obtained are still extremely complicated. Thus, because of the difficulties in finding a full analytical expression, this question has been solved by means of numerical integration. *Fig. 4* presents results of the calculations for three bundles of a different number of conducting fibres, assuming subsequently in the formula (9) the amplitude values (*A, 2A, 4A*). The calculated changes in the resultant charge on depolarization are approximately bell-shaped.

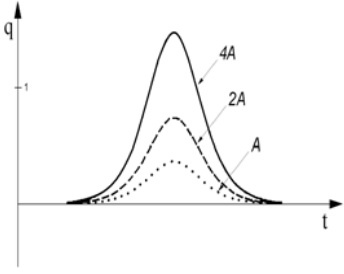

*Fig.4. Numerically calculated changes in the resultant electric charge ( q ) on depolarization of CM bundle cells of a different number of fibres (A, 2A and 4A)*

## 4. Electric potential

The dynamics of charge density distribution presented here is reflected in the potential, measured on the body surface, whose value at a given measuring site depends, among other things, on the value of the resultant electric charge at a given moment. The contribution to the electric potential coming from a single charge is of isotropic character. On the other hand, the potential measured is anisotropic. One should note however, that in separate heart regions, the processes discussed start at different moments in time. Moreover, each fibre bundle has a different orientation in space, which leads to an even more complicated potential distribution.

In general, the electric potential appearing around a single fibre is defined by a simple relation:

$$\varphi(\vec{r},t) = \int \frac{\Delta\rho(\vec{r}',t)}{|\vec{r}-\vec{r}'|} dv' \qquad (14)$$

where $\Delta\rho$ is the resultant charge density in the infinitesimal volume element *dv'*. The position of this element is determined by the vector $\vec{r}'$, shown in *Fig. 5*. Its length is a distance of the element volume *dv'* from the fibre centre, which is the origin of the coordinate system. Besides, the vector $\vec{r}$ describes a position of the measuring point in this system. According to the model discussed, each fibre studied is built of three parts (*W,M,N*) and each of them is a source of independent potential.

Considering interrelations between them, the total potential value can be calculated at any point *P* at any time of the depolarization process of the whole fibre.

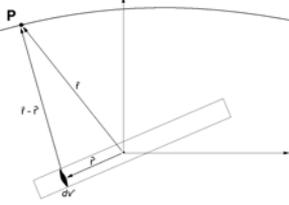

*Fig. 5.* Fragment of a CM fibre including infinitesimally small volume dv' with a defined coordinate system. P represents a measuring point.

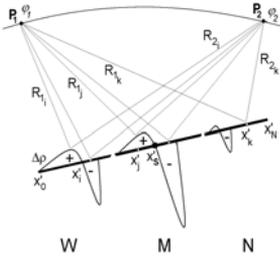

*Fig. 6.* The effect of charge density at individual parts of the fibre (WMN) on the value of the resultant potential φ at points $P_i$ (i = 1,...,n)

The scheme of the effect of the phenomena taking place in individual fibre parts on the value of the total potential is shown in *Fig. 6*. The figure shows the courses of the resultant charge density $\Delta\rho(x, t_k)$ calculated at a given time moment $t_k$ for each of the fibre parts.

Therefore, the potential at any point $P_i$ (i = 1,…,n) can be calculated by summation of all contributions to the resultant charge density distribution coming from each element of the fibre at the time *t*. A separate issue is to calculate the potential coming from the whole bundle, which is a sum of potentials produced at each of the excited fibres in it. Now, we have to take into account the position of each fibre in relation to the geometrical centre of the bundle, which is the origin of the coordinate system. Mathematical details of this operation will be left out, and in general, the potential appearing at point P on the bundle depolarization can be presented as:

$$\varphi_{(P)}(P,t) = \sum_i \varphi_i(P,t) \qquad (15)$$

where summation runs over all fibres of a given bundle. Moreover, at a given moment of time the activated *CM* bundles are at different stages of depolarization and bring their own independent contributions to the potential measured at a given point. Hence, the value measured on the body surface at *P* is a sum of all potentials existing at a given time. Then the resultant potential $\varphi_W$ can be expressed through the relation:

$$\varphi_W(P,t) = \sum_j \varphi_{(P)_j}(P,t) \qquad (16)$$

where the summation runs over all bundles taking part in the *CM* depolarization process (j = 1,…,10) taking into account to their position and orientation in a given coordinate system. The time *t* is measured during the whole depolarization process. The potential defined with the formula (16) comes from the charges from a limited region, assuming such a potential scaling in which they would disappear in an infinite distance. In practice on the body surface we measure the potential difference between that of a given electrode and the reference potential $\varphi_{ref}$, determined by measurement at the remaining electrodes. In the model presented we also take into regard the reference potential calculated from Einthoven's triangle according to analogous rules of the standard ECG examination. This potential is a sum of potentials calculated at the points of the limb electrodes in the standard ECG examination. The value of the reference potential also depends on the moment of time of the bundle depolarization process. Then the total potential $\varphi_C$, which is in fact a difference in the above-mentioned potentials ($\varphi_W, \varphi_{ref}$) is described by the following relation:

$$\varphi_C(P,t) = \varphi_W(P,t) - \varphi_{ref}(t). \qquad (17)$$

Thus, the potential value $\varphi_C$ depends on the position of the measuring point *P* (*x,y,z*) and on time in which the depolarization process is considered. Numerically calculated potential distributions $\varphi_C$ for a given *CM* bundle depending on the physiological and pathological factors will be discussed in detail further in this paper.

## 5. Material and methods

To verify the fundamental assumptions of the model proposed, a test study was performed on three groups of patients (*A,B,C*). Groups A and B consisted of healthy people, whereas in the group *C* there were patients suspected of cardiac muscle ischemia. The patients from group *C* were also subjected to the perfusion scintigraphy examination SPECT and stress ECG. The program of study was carried out according to methodology developed by Prof. Dr. hab of Medical Sciences Andrzej Dąbrowski after receiving a written approval of patients and an agreement of the Bioethics Committee of the Military Medical Chamber in Warsaw.

The main aim of examination of group A patients was to determine a range of changes in values characteristic of each of the instantaneous potentials and to calculate their interrelations. The values obtained served as a reference standard for verification of further results. On the basis of the examination of group B patients a spatial distribution of potential values on the thorax surface was calculated for each of the instantaneous excitations on *CM* depolarization. The results of examination of group *C* patients were used to determine usefulness of the *SATRO* method in prediction of the myocardium perfusion disturbances, diagnosed by the stress perfusion scintigraphy (SPECT). All patients underwent 12 lead ECG examinations whose results were subjected to *SATRO* analysis.

Group *A* consisted of 183 people at the age of 45±15. 100 people were in good physical condition (65 men and 35 women) with no symptoms of cardiac or cardiovascular system disorders or any risk factors of coronary heart diseases. The remaining 83 people (64 men and 19 women) were patients whose SPECT examination did not reveal any perfusion changes only in some areas of the left ventricle. The instantaneous potentials corresponding to these areas were taken into consideration in establishment of the standard.

Group B consisted of eight healthy men of a slim figure at the age 25-28. In a specially prepared room these men were subjected to measurement of electrocardiographic potentials with the use of 144 disposable electrodes placed on the whole chest surface. Four of these electrodes corresponded to the limb leads. The course of the potential changes was recorded subsequently from each electrode placed in the middle of the element of the net with side lengths of 35 mm horizontally and 50 mm vertically, as shown in *Fig. 13*. All the courses were analysed with the use of the *SATRO* programme, obtaining precise parameters for each of ten instantaneous excitations whose duration was calibrated on the basis of the corresponding potentials from the lead II.

The third group consisted of 86 people, 47 men and 19 women suspected of the cardiac muscle ischemia, qualified for stress ECG on the grounds of recommendations of classes I or II b, according to experts from the American College of Cardiology and American Heart Association [28, 29]. In the group, there were 20 men without any pain in the chest with one or two risk factors of the coronary heart disease. A clinical characteristic of this group was made and the type of ECG irregularities in the range of ST-T waves, classified according to the *Minnesota Code* [30] was determined. All persons from all groups were subjected to standard ECG and *SATRO*, SPECT and stress ECG.

### 5.1. *SATRO* analysis of ECG signals

12 lead ECG signals at rest were recorded on a standard apparatus permitting measurements in the frequency band corresponding to AHA norm for analog filtration (0.03 -100 Hz) and the sampling frequency of 500 Hz. The results were analysed by the program *SATRO* based on the physical model presented above. Signals from each lead were normalized and averaged according to the accepted criteria, and the isoelectric line (zero) was determined on the basis of the T-P section of the ECG records obtained.

The majority of biomedical signals, including those from ECG, are nonstationary so their course and preliminary parameters of each of the instantaneous potentials were estimated in the time-frequency signal representation. Analysis in this domain was possible as a result of a numerical operation additionally employing some algorithms permitting improvement of the resolution of this representation. A similar signal conversion procedure was described in detail in [31, 32] and the possibility of the resolution improvement was discussed in [33, 34]. The values of the parameters determining the potential magnitude were then optimised to get the best possible agreement between the course of the resultant function (thick solid line in *Fig. 7*) and the experimental data (•). Having taken into account the assumptions of the model presented in this work, the optimisation was performed by the two-phase method based on the genetic algorithms [35, 36]. *Fig. 7* presents the time courses of all potentials appearing on *CM* depolarization and only ten single-phase instantaneous potentials. Some of the potentials are two-phase in character, which illustrates the complexity of the QRS complex analysis by *SATRO*. As a result, the procedure gave the accurate values of the parameters of each of the instantaneous potentials (-) and their components (--) originating from particular parts of an activated bundle. As follows from the analysis of the QRS complex within the model proposed, upon depolarization of the cardiac ventricles a number of instantaneous activations take place. Each of them is related to a distribution of the resultant charge density $\Delta\rho_i$ in particular parts of a given bundle *(W,M,N)*.

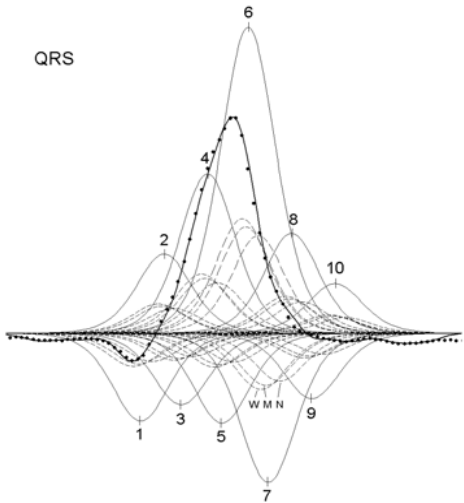

*Fig.7. All the potentials analysed by SATRO appearing on ventricular depolarization. Time dependencies of the total potential (-), experimental potential (•), instantaneous potentials (-) and the potentials originating from (W,M,N) parts of individual CM bundles.*

Actually, the majority of the potentials can be two-phase (*Fig. 8*), as indicated by the complicated character of analysis of the QRS complex by *SATRO*. For the sake of clarity the potentials from particular parts of the bundles are not marked.

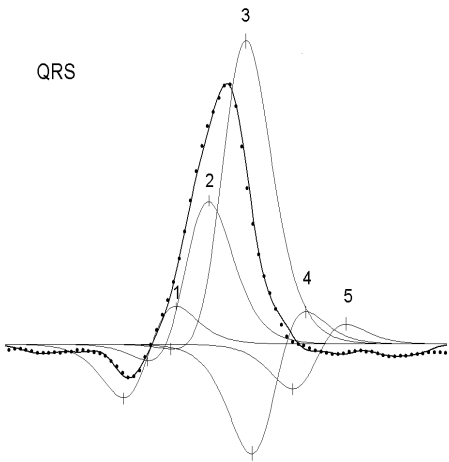

*Fig. 8. The two-phase potentials that can appear on the cardiac muscle depolarisation. The time dependencies of the total (-), experimental (•) and transient potentials (-) generated by individual bundle.*

Following the above procedure the signals from all leads of the standard ECG were analysed and the results were transformed into the orthogonal Franck's system on the basis of the empirical coefficients. According to the known methods of determination of the electric axis of the cardiac muscle [37, 38] the angles made to it by the physical axes of the cardiac muscle (saggital, transversal and anatomical) were estimated. At the next step the orthogonal system was transformed to the coordinate system based on the anatomical axes of the cardiac muscle so that the *y* axis corresponded to the long anatomical axis and the *z* axis to the saggital axis. The transformations employed the matrices used in the three-dimensional dextrorotatory system of the following directions of the positive rotation [39]:

$$x \rightarrow y, \quad y \rightarrow z, \quad z \rightarrow x \qquad (18)$$

In each case the coordinate system based on the anatomical axes was oriented with respect to the orthogonal system so that the resulting angles were:

- to the OX axis: -15.5°    (where "+" corresponds to Y -> Z)
- to the OY axis: -39.0°    (where "+" corresponds to X -> Z)
- to the OZ axis: -35.0°    (where "+" corresponds to X -> Y)

In order to evaluate changes in each potential generated on *CM* depolarization, a concept of instantaneous electric activity $V_i$ was introduced. This activity is measured as the area under the curve corresponding to a given instantaneous potential higher than $2 \times 10^{-6}$ [$V$], in the units of $10^{-6}$ [$Vs$]. The resultant activity $V_{ch}$ is calculated from eq. (19) and given in [%], where $V_i$ and $W_i$ are

$$V_{ch_i}[\%] = V_i \bullet 100\% / W_i \qquad (19)$$

the instantaneous electric activity for a given patient and the standard. However, more often the relative instantaneous activity $V_{zc}$, is used, calculated from relation (20), where $V_c$ and $W_c$ are the total activity for a given patient and the standard:

$$V_{zc}[\%] = (V_i / V_c) \bullet 100\% / (W_i / W_c) \qquad (20)$$

The final analysis of the relative instantaneous potential is carried out automatically according to our own procedure taking into regard the characteristic parameters of the four closest activations. Duration of each instantaneous potential is calculated relative to that of the corresponding potential in healthy patients, according to eq. (21) where: $t_{V_i}$, $t_{W_i}$ are the durations of each particular instantaneous potential

$$t_{ch_i}[\%] = t_{V_i} \bullet 100\% / t_{W_i} \qquad (21)$$

(i = 1÷10) of a magnitude higher than $2 \times 10^{-6}$ [V] for a given patient and the standard. The time interval to the next activation is calculated from $\Delta t_{i+1,i} = t_{i+1} - t_i$, where $t_i$ is the time at which a given activation potential reaches an extreme value.

The *SATRO* method presented in this paper is based on the above-described physical model and employs a numerical analysis of standard ECG signals. The obtained accurate parameters describing each of the instantaneous potentials are compared with the corresponding values measured for healthy patients, treated as standards. On the basis of comparisons of the disturbances in perfusion in individual fragments of the left ventricle determined in the SPECT with the changes in $V_{zc}$ calculated by the *SATRO* method a possibility was developed to estimate the performance of the following fragments of the left ventricle: septum (PR), anterior wall (ŚP), inferior wall (ŚD), lateral wall (ŚB), posterior (ŚT) and the apex (K). In the majority of the patients examined a decrease in the relative instantaneous electric activity in at least one anatomical area by over 50% relative to the corresponding standard was interpreted as a positive result of *SATRO*. Final evaluation of the cardiac muscle ischemia was made on the basis of this result analysed against other parameters describing this activation.

### 5.2. Myocardial perfusion scintigraphy SPECT

Stress ECG was performed on a moving track according to the Bruce protocol [40]. The heart beat frequency was established by following the rule: 220/min − age of the patient in years. The workload was ceased on a positive result or − in patients with the negative result − on reaching the maximum heart beat frequency or on detection of absolute or relative indications of test termination [28]. The criterion of a positive result of the stress test was the horizontal or a skew decrease or an increase in the ST interval ST ≥ mm in at least one of the 12 leads of stress ECG [28].
The rest-stress myocardial perfusion scintigraphy with tomographic imaging (SPECT) was made using methoxy-izobutyl-izonitrile with $Tc^{99m}$ isotope as a label. The examination was performed at the Nuclear Medicine Centre of the Army Medical Centre in Warsaw (headed by Prof. dr hab. med. Eugeniusz Dziuk) and at the Nuclear Medicine Department of the Institute of Cardiology in Warsaw (headed by Dr. med. Anna Teresińska). In analysis of the results of *SATRO*, SPECT and stress ECG the frequency of the positive results obtained by each method was established and then a positive result of SPECT was predicted on the basis of the results of *SATRO* and stress ECG and the prediction was verified. Moreover, the consistence of the SPECT and *SATRO* results was tested in the indication of perfusion deficiency localisation by SPECT and regions of diminished electric activity by *SATRO*.

The interrelations between the results of SPECT, *SATRO* and stress ECG were evaluated by the logistic regression test and Spearman correlation. Statistical analyses were made using the Complete Statistical Systems packet (Microsoft Corporation, USA). The sensitivity of *SATRO* and stress ECG was calculated as the percent of persons with positive results in the group of persons with perfusion disturbances diagnosed with SPECT. The sensitivity and specificity of *SATRO* and stress ECG was determined on the basis of a comparative analysis of the results obtained by these methods with those obtained by stress SPECT for the patients diagnosed with disturbances of the cardiac muscle perfusion. Moreover, the consistency of the results obtained by SPECT and *SATRO*, that is the agreement between the localisation of the perfusion deficiency regions detected by SPECT and the regions of diminished electric activity detected by *SATRO*, was evaluated.

As follows from the Spearman correlation test, the correlation of SPECT results with those of *SATRO* was stronger ($r = 0.46$, $p < 0.0001$) than with those of stress ECG ($r = 0.26$, $p < 0.02$). The sensitivity of *SATRO* in prediction of myocardial perfusion disturbances was 100 % and the *SATRO* specificity was 46%. After some correction of this range the sensitivity was 93 %, while the specificity - 81%. Values of these parameters for particular segments of the left ventricle are given below.

# 6. Results

This paper reports only some selected more important results needed to verify the assumptions of the model proposed. A detailed analysis of *SATRO* results and discussion of the advantages of its application in diagnosis of the myocardial ischemia and other cardiac muscle conditions will be presented in separate papers.

### 6.1. Standard parameters of instantaneous potentials

On the basis of the results obtained for patients from group A, the characteristic parameters of ten instantaneous potentials appearing on the chest surface as a result of electric activation of particular *CM* segments were calculated by the *SATRO* method. They bring independent contributions to the total potential measured by ECG.

*Table I* presents mean values of instantaneous potentials $\overline{W}_i$, mean total electric activity $W_c$ and the confidence intervals calculated.

*Table I. Mean values of 10 instantaneous electric potentials $\overline{W}_i$ and confidence intervals calculated for healthy patients.*

| Potential (i) | 1 | 2 | 3 | 4 | 5 | 6 | 7 | 8 | 9 | 10 | $W_c$ |
|---|---|---|---|---|---|---|---|---|---|---|---|
| $\overline{W}_i \times 10^{-6}$ [Vs] | 12.2 | 11.9 | 16.0 | 25.5 | 21.7 | 89.5 | 36.5 | 27.9 | 18.3 | 15.0 | 91.2 |
| Confidence interval $10^{-6}$ [V·s] | ±1.7 | ±1.6 | ±1.9 | ±2.8 | ±2.5 | ±10.6 | ±3.8 | ±3.3 | ±2.3 | ±1.5 | ±10.7 |

For each person examined the instantaneous potential at the sixth activation $W_6$ was significantly greater than the potentials corresponding to the other activations and equalled 89.5 x $10^{-6}$ [V·s] on average. The confidence intervals did not exceed 15 % of their values. *Table II* presents the durations of the activations and the calculated confidence intervals. The differences in the durations of activations are rather small, but the longest one of 52.3 x $10^{-3}$s characterises the sixth activation. The durations of the other activations fall within the range 38 ÷44 s.

*Table II. Mean duration of 10 instantaneous potentials and confidence intervals calculated for healthy persons.*

| Potential (i) | 1 | 2 | 3 | 4 | 5 | 6 | 7 | 8 | 9 | 10 | $T_c$ |
|---|---|---|---|---|---|---|---|---|---|---|---|
| $\overline{t}_i \times 10^{-3}$ [s] | 41.7 | 41.0 | 42.3 | 44.3 | 39.0 | 52.3 | 43.0 | 41.0 | 38.3 | 37.0 | 106 |
| Confidence interval $10^{-3}$ [s] | ±1.8 | ±1.8 | ±1.9 | ±2.1 | ±1.5 | ±2.3 | ±2.0 | ±1.9 | ±1.8 | ±1.6 | ±9.7 |

Analysis of time dependencies of particular activations permitted calculation of the time distances *Δt* between maximum values of instantaneous potentials of subsequent activations, see *Table III*.

*Table III.* *Mean time distance $\Delta t$ between the maximum values of instantaneous potentials of subsequent activations and the confidence intervals for healthy persons.*

| Activations<br>i ÷ (i+1) | 1÷2 | 2÷3 | 3÷4 | 4÷5 | 5÷6 | 6÷7 | 7÷8 | 8÷9 | 9÷10 |
|---|---|---|---|---|---|---|---|---|---|
| $\Delta t_{i+1,i}$ x $10^{-3}$ [s] | 7.3 | 7.3 | 8.0 | 6.0 | 8.0 | 8.0 | 7.0 | 7.7 | 7.0 |

The results obtained for healthy patients indicate that the time distances between the subsequent potential maxima are of the same value. Moreover, in all healthy persons examined the contributions of particular instantaneous activations in the total activation of the cardiac muscle were also of the same value. The parameters of subsequent potentials calculated on the basis of the results obtained for the healthy persons were treated as a reference standard. The reproducibility of results was also evaluated on the basis of the results of a series of 30 measurements performed in a few minute intervals for each of the 80 of healthy persons. The deviation from the mean value did not exceed ±10%.

### 6.2. Potential distribution on the chest surface

*Fig. 9* presents the time dependencies of the electric potentials measured on the chest surface of a healthy man of 25 years of age and of slim figure. Depending on the site of the electrode location, the potentials have different values and different courses. Each of them was analysed by *SATRO* to get the characteristic parameters describing the potentials. On the basis of the electrical activities calculated in this way it was possible to obtain a distribution of each of these potentials on the surface of the chest, see below. For the sake of illustration, *Fig. 9* presents the mean amplitude of one of the potential corresponding to the 6$^{th}$ activation of *CM*. The positive and negative values are marked in red and blue, respectively. On the front of the chest area there is one centre of the maximum potential amplitude of -1.60•$10^{-3}V$, localised between the leads V1 and V2, and another one below the lead V4 of the potential amplitude of +1.48•$10^{-3}V$. For the other patients from this group the potential distribution on the chest area did not differ significantly from that presented in *Fig. 9*.

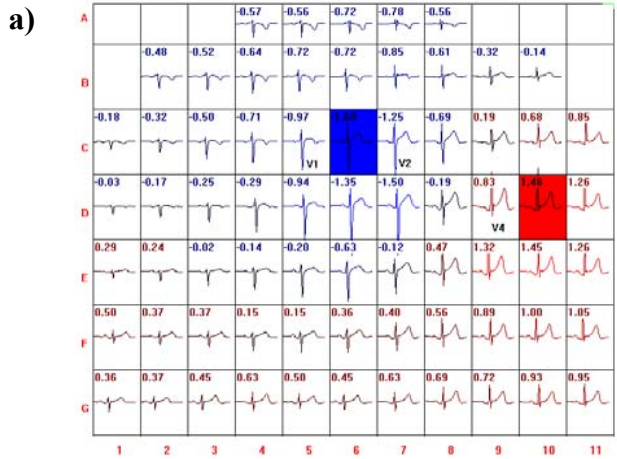

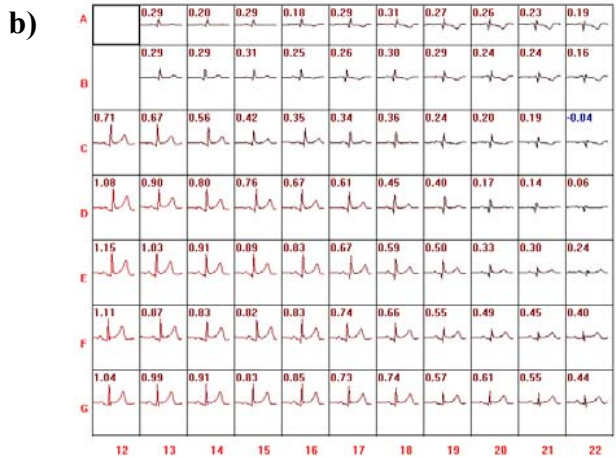

***Fig. 9.*** *Distribution of the amplitudes of the ECG signals measured: (**a**) on the front of the chest and (**b**) on the back, in a healthy man of 25 years of age. The values refer to the maximum amplitude of the $6^{th}$ activation expressed in $10^{-3}$ [V].*

## 6.3. Results of SPECT and *SATRO*

Analysis of the SPECT results by two independent groups proved the perfusion deficiency in at least one fragment of the left ventricle muscle in 65% of the patients examined. The positive *SATRO* result corresponding to a decrease in the local electric activity was detected in 78% of the patients from the same group. No person with the negative *SATRO* result was diagnosed with the perfusion deficiency in SPECT. The perfusion deficiency in SPECT and the reduced electric activity in *SATRO* were observed most often in the inferior wall and least often in the posterior wall of the cardiac muscle. A comparative analysis of the localisation of the perfusion deficiency in SPECT and reduced electric activity in *SATRO,* performed for a total number of 430 anatomical segments of the left ventricle proved the agreement of the results in 75% of the anatomical segments. The disturbances in the electrical activity measured by *SATRO* were assumed as significant when below 55% W(I) and below 50% W(II). The data on the sensitivity, specificity, positive (+) and negative (-) diagnostic values of *SATRO* are presented in *Table IV*.

**Table IV.** *Sensitivity and specificity of SATRO in prediction of perfusion deficiency in the selected anatomical segments of the left ventricle.*

| Sites at the left ventricle | W(I) | PR | ŚP | ŚD | ŚT | ŚB | W(II) |
|---|---|---|---|---|---|---|---|
| **Sensitivity (%)** | 100 | 71 | 45 | 88 | 50 | 100 | **93** |
| **Specificity (%)** | 48 | 68 | 88 | 69 | 89 | 73 | **81** |
| **Diagnostic value (+) (%)** | 48 | 29 | 56 | 77 | 9 | 33 | **90** |
| **Diagnostic value (-) (%)** | 100 | 93 | 84 | 83 | 99 | 100 | **86** |

## 7. Analysis and discussion of results

The hitherto known descriptions of the potentials whose appearance accompanies electric activation of ventricles have been based on the assumption that the electric dipoles making the front of the depolarization wave move along certain paths in the cardiac muscle. A detailed analysis of the process indicates that such an approach dos not give a coherent description of the physical phenomena taking place and is insufficient for evaluation of the parameters of particular activations. It also fails to take into account the effect of physiological and pathological factors on the space-time distribution of instantaneous potentials.

In the model presented the process of the cardiac muscle depolarization is accompanied by the movement of the resultant electric charge producing the potential, along the fibres. The problem has been analysed by *Clayton et al.,* [41] who studied the move of the potential along the single fibre. The distribution of each of the instantaneous potentials on the chest surface at a given time $t_o$ depends on the physiological and pathological factors. The effect of physiological factors on the potential was evaluated taking into regard different parameters related to the structure and position of the cardiac muscle. Particular anatomical segments of the cardiac muscle are represented by characteristic bundles. Their spatial position was described in the orthogonal system of coordinates (see *Fig. 11*). The angles $(\alpha, \beta)$ describe the bundle position in the system related directly to the patient and indicate the direction of propagation of the depolarization wave from the endocardium *W* towards the epicardium *N*. The distance between the centre of the bundle and the plane $(XY)_n$ tangent to the front of the chest is denoted by $r_o$.

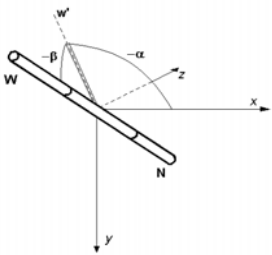

*Fig. 10. Position of the cardiac muscle bundle with its endocardium (W) and epicardium (N) in the orthogonal system of coordinates (xyz) fixed at the patient's body.*

The potential at a given site of the XY plane analysed depends on the distance $r_o$, angles $(\alpha, \beta)$ and the amplitude of the resultant charge density ($A_1$) of those at each active site of the bundle ($A_w$, $A_m$, $A_n$). Assuming $\alpha = 0°$ and $A_1$ = const., the values of the potential ($\varphi$) were calculated along the x axis for three different distances $r_o$ = 20, 30, 40. The plots are presented in *Fig.11a and 11b* for the angles $\beta = 0°$ and $\beta = -31°$, respectively. For $\beta = 0°$, for each of the three distances $r_o$, the potential course is asymmetric with the maximum value at the endothelium and the minimum at the epithelium. With increasing distance $r_o$ the absolute value of the potential decreases and the maximum and minimum get away from the centre of the bundle. The maximum of the potential is always closer to the centre and always from the side of the endothelium.

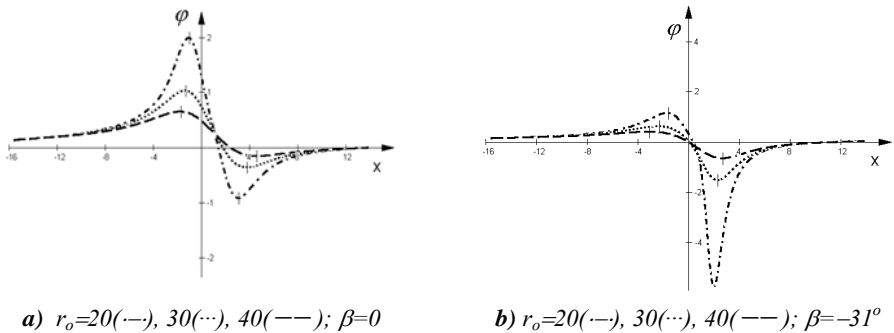

*a)* $r_o$=20(·–·), 30(···), 40(– –); $\beta$=0      *b)* $r_o$=20(·–·), 30(···), 40(– –); $\beta$=–31°

***Fig. 11.*** *The distribution of the potential $\varphi(x)$ for three different distances between the geometric centre of the bundle and the (XY) plane, for $A_1$= const. and the angles: **a)** $\beta$=0° and **b)** $\beta$ = –31°. The vertical signs indicate the extreme values of $\varphi(x)$.*

The rotation of the bundle in the YZ plane by the angle $\beta = -31°$, at the other parameters preserved, gives a much different distribution of the potential $\varphi(x)$, as shown in *Fig. 11b*.

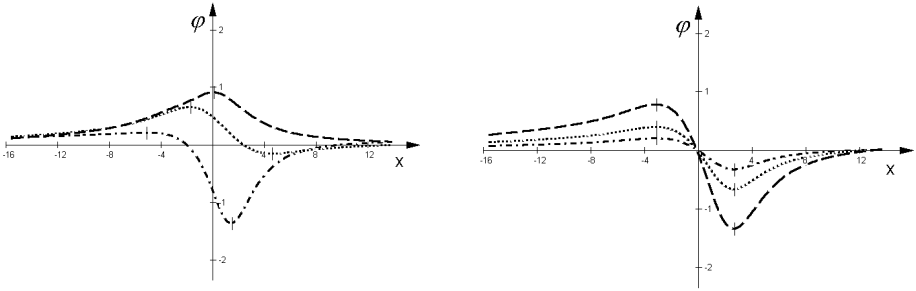

*a)* $r_o = 40$; $\beta = -57°(-\cdot-)$, $0°(\cdots)$, $57°(---)$   *b)* $r_o = 40$; $\beta = -31°$; $A_1(-\cdot-)$, $2A_1(\cdots)$, $4A_1(---)$

**Fig.12.** *The distribution of the potential $\varphi(x)$: **a)** for $r_o = 40$ and $A_1 =$ const. and for three different angles ($\beta$), **b)** for $r_o = 40$ and $\beta = -31°$ and for three different values of the amplitude $A_1$. The vertical signs indicate the extreme values of $\varphi(x)$.*

The absolute value of the potential for e.g. $r_o = 20$ is about three times higher than for the angle of 0. The potential distribution along the *x* axis at a distance $r_o = 40$ for the rotation angles $\beta = -57°$ and $\beta = +57°$ is of a completely different character as shown in *Fig. 12a*. Only one extreme of the potential appears (a positive or a negative one). In fact, individual bundles of *CM* are inclined at different angles and are at different distances from the measuring point, each brining its independent contribution to the resultant potential $\varphi$. Each bundle has a characteristic number of fibres, which has been taken into account in the calculations in the value of the amplitude $A_1$ related to the resultant charge density. *Fig. 12b* presents the distribution of the potential $\varphi(x)$ for three bundles of different *A*. As follows from the calculations, the magnitude of the amplitude does not affect the positions of the extreme values of the potential with respect to the centres of the bundles. The spatial distribution of the potential does not depend on the size of the heart for the same angle ($\beta$) and distance $r_o$.

On the basis of the assumptions of the model proposed, the distributions of the values of each of the 10 instantaneous potential were calculated on the front and on the back of the chest (*Fig. 13*). Particular distributions were determined at the time at which a given activation takes a maximum value. The calculations were performed for the characteristic parameters of the bundles, i.e. the angles $(\alpha, \beta)$, distances $r_o$ and the amplitudes $A_G$. The results are presented in *Fig. 13* for 10 instantaneous potentials. The upper part of each figure corresponds to the experimental results –marked by the letter *E*, while the lower to the theoretical results – marked by the letter *T*. All the potentials were recorded and calculated for one man from group B. The potential $\varphi$ distributions are presented in the form of the equipotential lines with decreasing intensity of the colour, the extremes are always marked with the same colour, to compare the values see the scales.

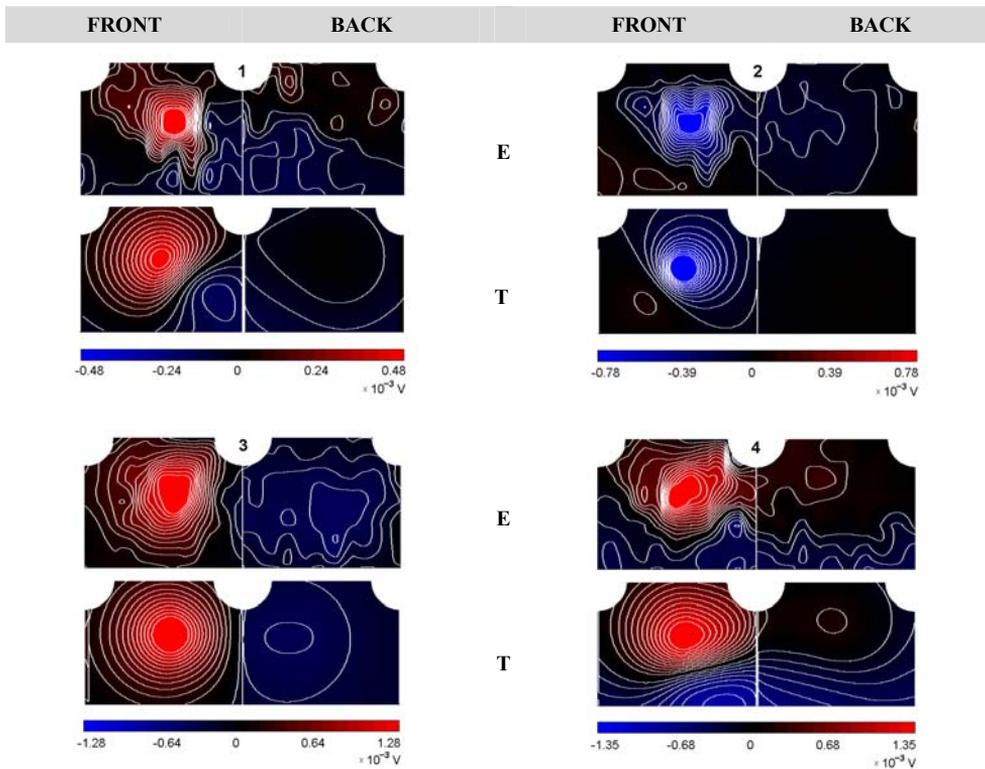

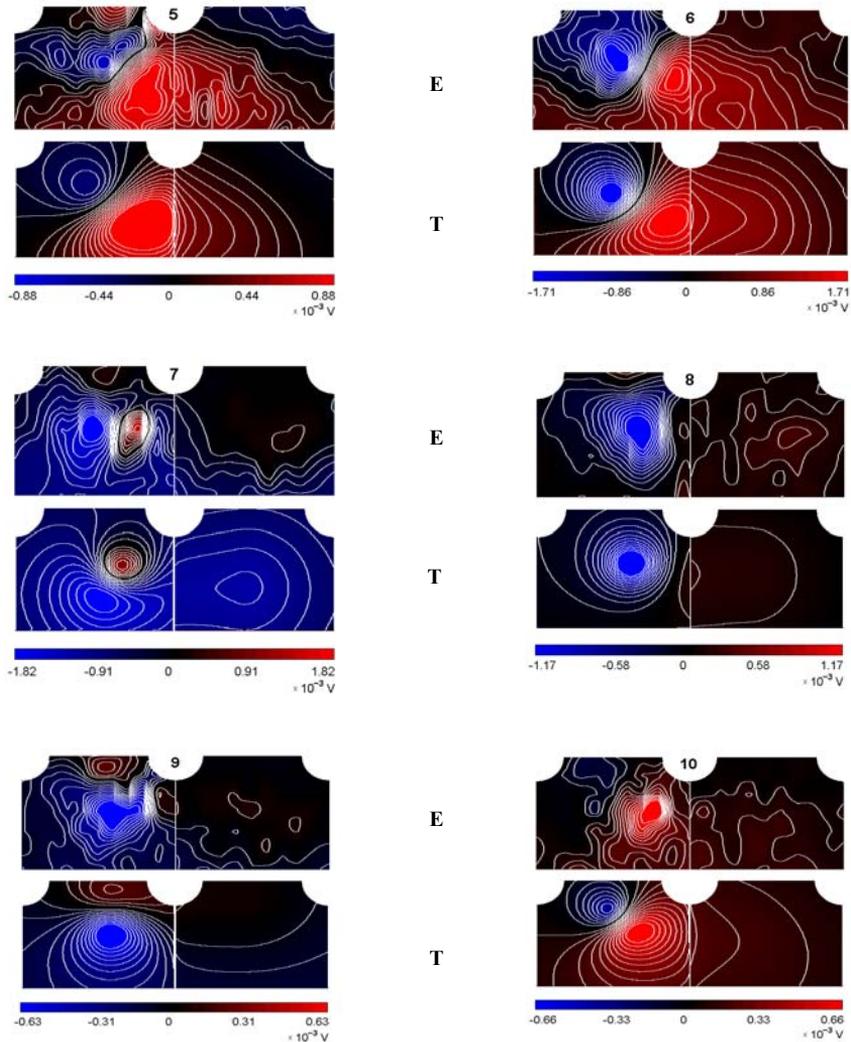

***Fig.13.*** *The distribution of the potential and the equipotential lines on the front surface of the chest for 10 instantaneous activations on depolarization of the cardiac muscle. The experimental data (E) are those collected for the patient from group B, while the theoretical data (T) were calculated on the basis of the model proposed.*

The number of points on the plot was increased by the Langrange interpolation of the potential measured, whose nodes were the potential values measured, which permitted drawing the plot to an accuracy of 1mm. To clarify the illustration the same colour intensity was used at the maximum value of each activation potential. The interrelations between the potentials can be estimated against a scale given below each figure. The insignificant differences between the potential distribution calculated on the basis of the model proposed and that obtained from measurements can be explained by assuming a linear structure of the *CM* bundles in the model. In fact some bundles can be somewhat curved.

Hitherto we have considered the spatial distribution of instantaneous potentials and at this point we shall focus on their time evolution. Assuming the model proposed an analysis has been made of time dependencies of each of the instantaneous potentials at a given site. Recently much attention has been devoted to the differences in the shape, duration and electro-physiological properties of the action potentials produced by the working fibres of particular segments of the ventricles [24, 25, 42]. Results of the hitherto studies have shown that the activation of cells in the subendocardial (*W*) and the *M* cells starts almost at the same time, which is probably related to the penetration of a certain number of Purkinje fibres deep under the endocardium [43]. The activation of the cells in the subepicardium is a little delayed. The duration of the potential at the activated *M* cells is slightly longer than at the other bundle parts because these cells make almost 40% of the whole bundle [44]. Following the above time relation, *Fig. 14a* presents the time evolution of the resultant potential $\varphi(t)$ of one of the *CM* bundles, calculated at the site $P(x,y,z)$, represented by the thick solid line (——). The potential time evolution $\varphi(t)$ is mainly affected by the properties of the particular parts W, M and N, marked by different type curves (---,——,•••,). The time $t_o$, at which the potential $\varphi(t)$ reaches a maximum value, is marked by a vertical line. The area *(P)* under the curve $\varphi(t)$ in *Fig. 14a* was calculated and assumed as a 100% in further analysis. Subsequent figures (*Fig.14b, 14c and 14d*) show the effect of the activities of particular parts of the bundle (*W, M, N*) on the value of *P*, $t_o$ time and the character of the $\varphi(t)$ course. For example, the neglect of the potential produced at the subendocardial part *(W)* of a given segment of *CM* (*Fig. 14b*) results in a decrease in the area under the f(t) curve *P* by about 40%, and the time of reaching the maximum potential values is by about $2 \times 10^{-3}$ s longer. If the *M* type cells do not take part in the depolarization (*Fig. 14c*), the area *P* decreases by about 34 % and the time $t_o$ remains the same.

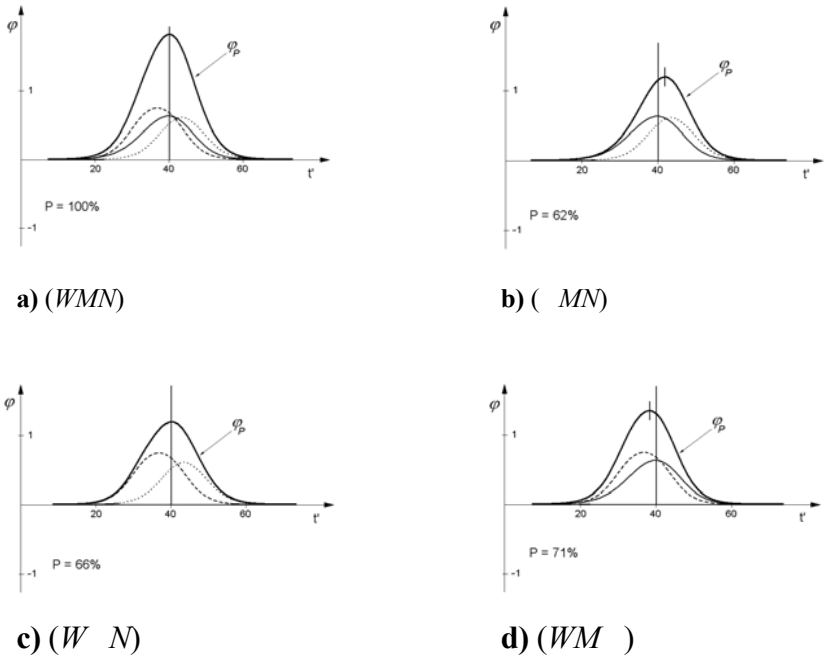

***Fig.14.*** *The time dependence of the resultant potential $\varphi_P$ produced at the CM bundle obtained taking into account: **a**) the subendocardial part W(---), the M type part M (—) and the subepicardial part N (•••), **b**) for the parts (M,N), **c**) for the parts (W,N), **d**) the parts (W,M).  **P** – is the percent of the area under each subsequent curve relative to that in Fig. 15a.*

The values of the instantaneous potentials and times after which they reach maximums depend on the contribution of particular parts of the bundle in the activation process. Within the model proposed it is also possible to analyse different combinations of contributions of each of the in parts, including the effect of the amplitude of partial potentials on the value of the instantaneous potential $\varphi_P$ (*Fig. 15a*) and that of the rate of the charge displacement in both areas I and II in each part of the bundle. If the difference in the rate of the ion movements in the regions I and II is significant, the dependence $\varphi(t)$ can also have a two-phase character (see *Fig. 15b*).

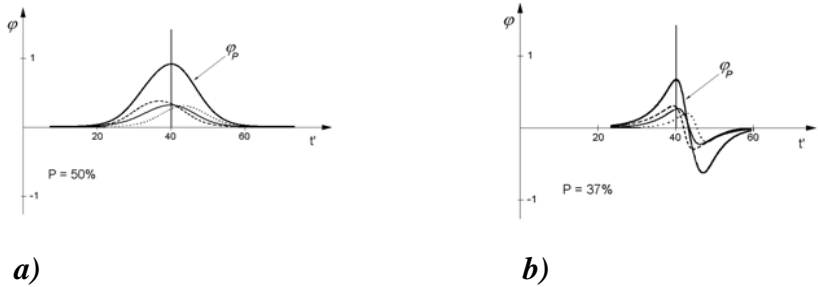

      *a)*                                 *b)*

**Fig.15**. *The time dependence of the potential $\varphi_P$ appearing on depolarization of the three parts of the CM bundle: **a**) of the amplitude ½A and **b**) for a high rate of the charge propagation in area II.*

Actually, the probability of such a situation is very low, but its occurrence cannot be excluded. On the basis of the model presented, the time courses of the instantaneous potential appearing around the bundle of fibres can be of one- or two-phase character, depending on the position relative to the centre of the bundle. The time courses are shown in *Fig. 16*. A detail analysis of the instantaneous potentials by *SATRO* is made taking into account all the possibilities that can occur simultaneously and bring different contributions.

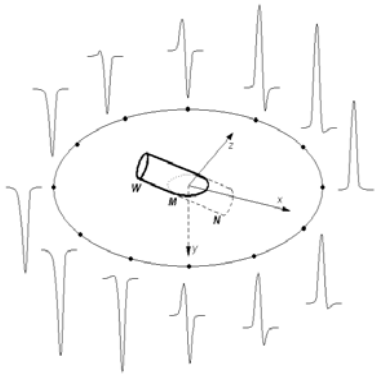

**Fig. 16.** *Time dependence of the instantaneous potential appearing around the bundle of fibres at sites equidistant from the centre of the bundle*

Despite the fact that in the model proposed the bundle is assumed to be made of fibres activated at the same time, each fibre brings its own independent contribution to the bundle performance and thus to the performance of a given segment of the cardiac muscle. Individual *CM* bundles are activated at different time moments and bring their own contributions to the potential appearing around the heart in a strictly determined sequence. Different parameters of each bundle and their mutual spatial arrangement implied by the structure of the cardiac muscle give rise to some anisotropic changes in the physical quantities at the chest area.

A comprehensive discussion needs to take into regard the dielectric properties of the human body on the electric potential measured.

In view of the fact that in the model proposed we use the arbitrary values and units and that the dielectric characteristics of the medium – a human body - for the real values of the parameters measured can be assumed as linear, the above equations are sufficient for our purposes. Nevertheless, it should be remembered that although these relations are not complicated, it is practically impossible to find an exact analytical expression for the potential, despite the assumed one-dimensional character of the model of each individual fibre. This impossibility is implied by the real form of the functions describing the densities of the positive and negative potentials and the geometry of the system. The problem is the form of the integrals describing the potential as a function of charge density of the fibres. Moreover, the total potential reflected by the signal measured is a sum of potentials from the resultant charges in individual fibres. Taking all this into account would require finding a compact analytical solution of the expression written for the limits of integrations determined by the positions of the beginnings and ends of each fibre and the summation should be performed over all fibres from the area of interest. Moreover, for each fibre we will have to find an appropriate integral taking into account that each fibre has a little different geometry and different time of the processes taking place in it, so the integration limits should be different for each fibre. However, it is possible to calculate numerically the potentials and hence the parameters of interest to us, to a satisfactory accuracy.

      The model proposed describes not only electric but also magnetic properties. Because of the numerical character of the method *SATRO* it is also possible to investigate the distribution of magnetic fields accompanying the activations. The problem is very interesting; however, it is beyond the scope of this work and will be a subject of a separate paper.

The results obtained by *SATRO* for the patients from group *A* have revealed significant differences in the electric activity in particular *CM* segments at small differences in durations of the activations. The appearances of the minimum or the maximum values of the instantaneous potential also occur at the same time intervals. This observation would suggest that the electric activity of particular segments of the ventricles comes from the activated *CM* bundles of a different number of fibres and slightly different lengths. The inferior wall is the only segment of the left ventricle represented by a single although the thickest bundle, which may be related to the occurrence of great number of infarcts in this segment.

Electrocardiographic diagnostics of ischemia or hypoxia of the cardiac muscle usually takes into account the localisation of the ST segments and the shape of the T wave. A characteristic feature indicating ischemia is a depression of the ST segments, while hypoxia is reflected by elevation of this segment, but the latter has lower diagnostic value. Each of the features is relatively little sensitive and little specific manifestation of ischemia, in particular if the ECG was recorded in the period when the patient suffered no pain and belonged to a group of persons in which the probability of coronary arteries disease estimated on the basis of the sex and age was low.

Small diagnostic value of the ECG components related to the repolarization of the muscles of the ventricles prompts to consider the possibility of using the process of the ventricle depolarization for the purpose. In the ischemic region of the cardiac muscle the content of ATP and creatine phosphokinase decreases, while the concentration of potassium, lactates and fatty acids in the extracellular space increases [45-47]. An increased concentration of extracellular potassium has an important effect on the electrophysiological properties of the ischemic muscle fibres. According to the experimental evidence, the accumulation of potassium ions caused by ischemia decreases the absolute value of the potential at rest and the rate of phase 0 increases together with the amplitude of the action potential [47,48], and decreases the values of electric potentials generated in the depolarization phase of the ischemic region of the ventricular muscle.

The disturbances in the process of depolarization related to ischemia are reflected in the shape of the QRS complexes. However, when the region affected by ischemia is small, this effect can be difficult or impossible to detect by standard evaluation and interpretation of ECG recording. Hitherto much attention has been paid to the broadening of the QRS complexes as a result of ischemia. Michaelides et al. [49] reported a correlation between the degree of broadening of the QRS complexes in stress ECG and the number of narrowed coronary arteries and the number of segments of the felt ventricle manifesting disturbances in contractility in isotope ventriculography. According to the authors, the range of broadening of the QRS complex as a result of ischemia is relatively small and the changes can be difficult on visual inspection of the electrocardiogram. For instance in the group of 150 persons examined by Michaelides et al., [49] the QRS complex broadening on stress was on average $4.8 \times 10^{-3}$s in the patients with significant narrowing of one coronary artery, $7.8 \times 10^{-3}s$ in the patients with narrowing of two coronary arteries and $13.3 \times 10^{-3}$s in the patients with narrowing of three coronary arteries. Such an insignificant broadening of the QRS complex can be explained by a decrease in the potential of some instantaneous activations, because, as follows from the model presented, hypoxia leads to a shortening of the activations and not to their elongation. The cardiac muscle cells contain a large number of mitochondria, so almost all energy is released as a result of oxidation of nutrients and the anaerobic processes have no importance. The deficiency of oxygen causes a delay of the charge transfer along the fibres in area II, which reduces the resultant charge density and thus also the potential measured. However, this has no effect on the propagation rate of the resultant wave of the charge density wave. The situation is different when one of the parts of the bundle does not take part in depolarization. Then, the effective length of the bundle is shortened, hence, the time of the activation and the value of the potential generated also decrease. Therefore, there is a strict relation between the ischemia of a given segment and the value of the resultant potential and duration of a relevant activation.

The evidence of depolarization disturbances related to hyperpotasemia in the region with ischemia reported by different researchers [47, 48, 50, 51] supports the correctness of the theoretical assumptions of the *SATRO* method. The proposed analysis of instantaneous variation of the depolarization potentials at particular sites in the left ventricle muscle extends the diagnostic possibilities of the non-invasive methods used for diagnosis of the cardiac muscle ischemia. A comparative study of the results obtained by the proposed method *SATRO*, stress ECG and SPECT performed on the whole group of patients studied, has shown a much greater correlation between the *SATRO* and SPECT results than between the stress ECG and SPECT. Such a good agreement between the results of *SATRO* and SPECT follows from the fact that both methods bring the information on the state of *CM*, and effects of the disturbances in intraventricular conduction and other diseases leading to local damage to muscle fibres, Purkinje fibres or Purkinje junction on the parameters analysed in both methods are the same. For this reason, the results obtained by the method *SATRO* and stress ECG were referred to the stress SPECT results and not to those of the angiography of coronary arteries.

## 8. Concluding remarks

Investigation of instantaneous potentials, appearing on the chest surface on depolarisation of the cardiac muscle, required development of a new model describing generation of the resultant electric charge density and new software permitting detailed analysis of the ECG signals. Changes in the charge density were described by mathematical formulae and the resultant distribution of potentials was then compared to the experimental results.

The model assumed describes in a coherent way the space-time relations between the potentials appearing on the cardiac muscle depolarisation taking into account the pathological and physiological factors.

Analysis of the ECG records by the *SATRO* program enables a detection of even small changes in the instantaneous potentials appearing as a result of activation of a given fragment of the cardiac muscle. This potentiality indicates a possible application of the method proposed for detection and evaluation of early ischemia in individual fragments of the cardiac muscle.

As has been evidenced, the diagnostic worth of *SATRO* in prediction of disturbances in the myocardial perfusion diagnosed by SPECT is much greater than that of classical ECG.

The conclusions following from the model and the method proposed, its analysis and tests of its diagnostic worth are given below:

1. The appearance of potentials at particular fragments of the cardiac muscle can be explained by analysis of changes in the electric charge in three-component (W, M, N) bundle of conducting fibres.
2. Each bundle represents a separate anatomical fragment of the cardiac muscle (ventricular septum, anterior wall, posterior wall, inferior wall, lateral wall, etc.) and is an independent source of instantaneous potential.
3. The QRS complex appears as a result of summation of all instantaneous potentials generated on depolarisation of particular fragments of the cardiac muscle.
4. The value of instantaneous potential depends on the properties and activities of individual parts (W, M, N) of a bundle.
5. Time changes in the value and distribution of the potential appearing on the surface of the chest depends on physiological and pathological factors.
6. Electric activity of each particular fragment of the cardiac muscle of the healthy person makes a constant percent contribution in the total cardiac muscle activity.
7. Particular activities of a value lower than 50 % of the healthy person (standard) well correlate with disturbances in perfusion observed in stress SPECT.
8. A reduced amount of oxygen in the cardiac muscle significantly affects the resultant electric charge density and leads to a reduced electric potential generated on the cardiac muscle activity.
9. The diagnostic worth of *SATRO* examination in predicting disturbances in the cardiac muscle perfusion diagnosed with stress SPECT is greater than that of stress ECG.
10. High sensitivity of *SATRO* in prediction of the positive result of stress SPECT indicates a possibility of its application as a screening test in detection of the cardiac muscle ischemia.

## Acknowledgements

My heartfelt thanks go to the following persons.

Prof. Ryszard Piotrowicz for coordination of the reference study of *SATRO* and for the opportunity of presenting the method at the International Conference of Non-invasive Electrocardiography of the Polish Cardiological Association.

Dr habil. Rafał Baranowski for kind help and valuable comments.

Prof. Andrzej Dąbrowski for working out the methodology of the research project and fruitful discussions.

Mr Mirosław Chąpiński and Mr Łukasz Janicki for editing this paper, computer simulations and computations.

Dr Anna Teresińska from the Department of Nuclear Medicine of the Institute of Cardiology in Warsaw, and

Prof. Eugeniusz Dziuk and Dr Mirosław Dziuk from the Department of Nuclear Medicine, Military Institute of the Health Services in Warsaw, for realisation of the reference studies.

Prof. Zbigniew Religa, head of the Cardiology Institute in Warsaw, for permission for the clinical study and the members of the Board of Directors of the Polish Cardiological Association for support of this study.

Mr Andrzej Ruszczak, president of the Foundation HEALTH AND CULTURE (ARAK) for promoting the *SATRO* program in Poland and valuable consultations.

Mrs Beata Janicka, my beloved wife and my two sons Łukasz and Michał for help, support and believe in success of my work, which meant a lot to me.

The project *SATRO* has been financed by the means of the Medical Physics Research Institute with some support in the range of scientific expertise from the EU funds.